\documentstyle[11pt,aasms4]{article}

\received{31 Jan 2000}
\accepted{30 Mar 2000}

\def\etal{et al.}
\def\Pdot{$\dot P$}

\slugcomment{Submitted to Ap.J.Letters}
\lefthead{Kepler et al.}
\righthead{Evolutionary timescale of G117-B15A}
\begin{document}

\title{Evolutionary Timescale of the DAV G117-B15A: The Most Stable Optical 
Clock Known}
\author{S.O. Kepler}
\affil{Instituto de F\'{\i}sica da UFRGS, 91501-900 Porto Alegre, RS - Brazil,
kepler@if.ufrgs.br}
\author{
Anjum Mukadam, D.E. Winget,
R.E. Nather, \& T.S. Metcalfe}
\affil{Department of Astronomy, University of Texas, Austin, TX 78712 - USA}
\author{M.D. Reed \& S.D. Kawaler}
\affil{Department of Physics and Astronomy, Iowa State University, Ames, Iowa - USA}
\author{Paul A. Bradley}
\affil{Los Alamos National Laboratory}

\begin{abstract}
We observe G~117--B15A,
the most precise optical clock known,  to measure the
rate of change of the main pulsation period of this blue-edge DAV white dwarf.
Even though the obtained value is only within 1~$\sigma$,
$\dot P = (2.3 \pm 1.4) \times 10^{-15} \,\rm s/s$, it is already
constraining the evolutionary timescale of this cooling white dwarf star.
\end{abstract}

\keywords{Stars: evolution -- stars:
oscillations -- stars: individual: G~117-B15A}

\section{Introduction}
We report our continuing study
of the star G~117--B15A, also called RY LMi, and WD0921+354,
one of the hottest of the pulsating
white dwarfs with
hydrogen atmospheres,
the DAV or ZZ~Ceti
stars (McGraw 1979).
McGraw {\&} Robinson (1976)
found the star was variable,
and Kepler \etal\ (1982)
studied its light curve and found 6 pulsation modes. The dominant mode is
at 215~s, has a fractional amplitude of 22 mma,
and is stable in amplitude
and phase.
The other, smaller pulsation modes vary in amplitude
from night to night (Kepler \etal\ 1995).
Because the DAVs appear to be normal stars except for their
variability (Robinson 1979, Bergeron {\etal} 1995),
it is likely that the DAV structural properties are representative
of {\it all}  DA white dwarfs.

The rate of change of a pulsation period
is directly related to the evolutionary timescale of a white dwarf,
allowing us to directly infer the age of a white dwarf since its
formation.
We have been working since 1975 to measure the
rate of period change with time ($\equiv \dot P$)
for the $P=215$~s periodicity
of G117--B15A, and the Kepler et al. (1991) determination was
$\dot P= (12.0 \pm 3.5) \times 10^{-15}\,{\rm s/s}$, including
all data obtained from 1975 through 1990.

Kepler (1984) demonstrated that the observed
variations in the light curve of G~117--B15A 
are due to non-radial {\it g}-mode pulsations
and therefore the timescale for  period change is directly proportional to
the cooling timescale.

For comparison, the most stable atomic clocks have rates of period
change of the order of $\dot P \simeq 2 \times 10^{-14}$~s/s,
while the most precise millisecond pulsars have
$\dot P \simeq 10^{-20}$~s/s 
(Kaspi, Taylor \& Ryba 1994 and references therein).
Since the stability of a clock is measured by
$P/\dot P$, G117--B15A has the same order of
stability as the most stable millisecond pulsar.

G117--B15A is the first pulsating white dwarf to have its main
pulsation mode index identified. The 215~s mode is an $\ell=1$, 
as determined by comparing the
ultraviolet pulsation amplitude (measured with the Hubble
Space Telescope) to the optical amplitude (Robinson \etal\ 1995).
Robinson \etal\ (1995), and Koester, Allard \& Vauclair (1994)
derive $T_{\mathrm{eff}}$ near 12,400~K, while Bergeron \etal\ (1995),
using a less efficient model for convection, derives $T_{\mathrm{eff}}$=11,600~K.

Bradley (1996) used the mode identification and the 
observed periods of the 3 largest known
pulsation modes to derive a hydrogen layer mass lower limit
of $10^{-6}\,M_*$, and a best estimate of $1.5 \times 10^{-4}\,M_*$,
assuming $k=2$ for the 215~s mode, and 20:80 C/O core mass. The
core composition is constrained mainly by
the presence of the small 304~s pulsation.

\section{Observations}

We obtained 19.6~h of time series photometry in Dec 1996
and Feb 1997, plus
18.8~h in Mar and Dec 1999,
using
the three-star (Kleinman, Nather \& Phillips 1996)  photometer on the $2.1$~m
Struve telescope at McDonald Observatory.

To maximize the signal-to-noise (S/N) we 
observed unfiltered light, because the nonradial
{\it g}-mode light variations have the same phase in all colors
(Robinson, Kepler \& Nather 1982).
G117--B15A has V=15.52 (Eggen \& Greenstein 1965).

\section{Data Reduction}
We reduce and analyze the data in the manner described by Nather \etal\
(1990), and Kepler (1993).
We bring all the data to the same fractional amplitude scale,
and transform the observatories'
UTC times to the uniform Barycentric Coordinate Time 
(TCB) scale (Standish 1998),
using JPL DE96 ephemeris
as our basic solar system model (Stumpff 1980).
Kaspi, Taylor \& Ryba (1994) show that the effects of using different
atomic timescales and ephemeris are negligible.
We compute Fourier transforms for each individual run,
and verify that the  main pulsation at 215~s
dominates each data set and has a stable amplitude.

\section{Time Scale for Period Change}

As the dominant pulsation mode at P=215~s has a stable frequency and amplitude
since our first observations in 1975, we can calculate the
time of maximum for each new run and look for deviations due to
evolutionary cooling.

We fit our observed time of maximum light to the equation:
$$(O-C) = \Delta E_0 + \Delta P \cdot E
+ {1 \over 2} P \cdot \dot P \cdot E^2,$$
where $\Delta E_0 = (T_{max}^0 - T_{max}^1)$, $\Delta P = (P - 
P_{t=T_{max}^0})$, and $E$ is the epoch of the time of maximum, i.e, the number 
of cycles
after our first observation.

In Figure~1, we show the O--C timings 
after subtracting the correction to period and epoch,
and our best fit curve through the data.
>From our data through 1999, we obtain a new value for the epoch of maximum,
$T_{max}^0 = 244\,2397.917509 \,{\rm TCB} \pm 0.5 \,\rm s$,
a new value for the period, $P = 215.197 390 7 \pm 0.000 000 6 \,\rm s$,
and most importantly, a rate of period change of:
\[\dot P = (2.3 \pm 1.4) \times 10^{-15} \,\rm s/s.\]

We use linear least squares to make our fit, with each point weighted inversely
proportional to the uncertainty in the time of maxima for each 
individual run squared.
We quadratically add an additional $1.8$~s of uncertainty to the time of maxima
for
each night to account for external uncertainty caused perhaps
by the beating of small amplitude pulsations (Kepler \etal\ 1995)
or small amplitude modulation.

The estimated $\dot P$ is substantially different from the value estimated
in 1991. The apparent reason is a scatter of the order of 1.8~s present
in the measured times of maxima. Kepler et al. (1995)
discuss the possibility of such scatter being caused
by modulation due to nearby frequencies, and
Costa et al. (1999) shows that the real uncertainties 
must include the effect of all periodicities present.  
The 1991 value did not include
such scatter in the uncertainty estimation, and resulted in an 
overestimated statistical accuracy. We now treat this scatter
as an external source of noise.

\section{Core Composition}

For a given mass and internal temperature distribution,
theoretical models show that the rate of period change
increases if the mean atomic weight of the core is increased,
for models which have not yet crystallized in their interiors.
This applies to G117--B15A, as it is not cool enough to have a 
crystallized core (Winget et al. 1997).
Bradley, Winget {\&} Wood (1992) and Bradley (1998) compute rates
of period change for models that are applicable to G117--B15A, and we
summarize the relevant results here.
The models of Bradley, Winget, {\&} Wood (1992)
and Bradley (1998) are full evolutionary models that include compositional
stratification, accurate physics, and use the most recent neutrino emission
rates. We refer the reader to Bradley, Winget, {\&} Wood (1992) and Bradley
(1996, 1998) for further details.

Two major known processes govern the rate of period change in the
theoretical models of the ZZ Ceti stars:
residual gravitational contraction, which causes the periods to become
shorter, and cooling of the star, which increases the period as a result of the
increasing degeneracy (Winget, Hansen, {\&} Van Horn 1983),
given by
\[\frac{d(\ln P}{dt} = -a \frac{d\ln T_c}{dt}+  b \frac{d\ln R}{dt}\]
where $a$ and $b$
are constants associated with the rate of cooling and contraction
respectively, and are of order unity.

Following Kawaler, Hansen, {\&} Winget (1985), 
we can write 
\[\frac{d\ln P}{dt} = (-a + bs) \frac{d\ln T_c}{dt}\]
where $s$ is the ratio of the contraction rate to the cooling rate
\[ s \frac{d\ln T_c}{dt} = \frac{d\ln R}{dt}\]
or
\[ s = \frac{d\ln R}{d \ln T_c}.\]
The $dt$ terms cancel because we evaluate the derivative as the differences
in the radius, core temperature, and age between two models.
Spectroscopic $\log g$ values suggest that
G117-B15A has a mass between 0.53~$M_\odot$ (Koester \& Allard 2000) and
0.59~$M_\odot$ (Bergeron et al. 1995), and this agrees with the
preferred seismological mass range of 0.55 to 0.60~$M_\odot$ (Bradley 1998).
For a DA white dwarf near 12,000 K, the radius is about $9.6\times 10^8$~cm,
with a contraction rate of about 1~cm~yr$^{-1}$. The core temperature 
is
about $1.2\times 10^7$~K, with a cooling rate of about $0.05$~K~yr$^{-1}$.
With these numbers, $s$ is about $0.025$, which confirms our expectation that
the rate of period change is dominated by cooling. Other processes, such as
rotational spin-down and magnetic fields must be small, because we do not
see reliable evidence of either in 
the fine structure splitting of the observed
frequencies.

Bradley (1998) give a $\dot P$ value of 
$3.7   \times 10^{-15}\,\rm s/s,$ and find a spread of
$\pm 1 \times 10^{-15}\,\rm s/s,$ predicted by the range of acceptable 
models for G117--B15A, with the $0.60 M_\odot$ models having the smaller
values. 
His predicted value is within the $1\sigma$ error bars of 
the observed value; a more precise observational $\dot P$ determination
could in principle suggest a favored stellar mass.

Bradley's (1998) models are typically about 80{\%} oxygen, and 
Bradley~et~al. (1992) describe in detail the effect of changing the core
composition from pure carbon to pure oxygen for 0.5 and 0.60~$M_\odot$
models. 
They also show that the predicted $\dot P$ value from an 
oxygen core model is about 15 to 20{\%} larger than for an equivalent
carbon core model, rather than the 33{\%} predicted by Mestel (1952)
cooling theory.
This reduction in $\dot P$ from Mestel theory is the result of the ions 
being a Coulomb liquid, rather than an ideal gas as assumed by Mestel 
theory.

The $\dot P$ values quoted above are for the case where the 215~s mode
is not trapped (see Bradley 1996 for details), and Bradley~et~al. (1992)
show that if the 215~s mode is trapped, then the predicted $\dot P$ 
value could be as little as half the values predicted by Bradley (1998)
and quoted above.
In recent years, the $\dot P$ determinations have fluctuated
between about 1 and $3 \times 10^{-15}\,\rm s/s$ (see Table 1), 
so the values predicted
by seismological models are still consistent with the observations.
Reducing the observational errors to about half the present value of 
$1.4 \times 10^{-15}\,\rm s/s$ would provide enough of a constraint to 
confront the model predictions.

\section{Reflex Motion}
The presence of an orbital companion
could contribute to the period change we have detected.
When a star has an orbital companion, the variation of its line-of-sight
position with time produces a variation in the time of arrival of
the pulsation maxima, by changing the light travel time between the star
and the observer.
Kepler et al. (1991) calculated the possible contribution to $\dot P$ caused by
reflex orbital motion of 
the observed proper motion companion of G117--B15A as
$\dot P \leq 1.9 \times 10^{-15}\,\rm s/s$. 
If the orbit is highly eccentric and G117--B15A is near periastron, 
the orbital velocity could not be higher than twice that derived above
or it would exceed escape velocity. 
The above derivation
assumed that 
the orbit is nearly edge on to give the largest effect possible.
Therefore, $\dot P_{orb} \leq 3.8 \times 10^{-15}$ s/s.

The upper limit to the rate of period change could also be expected
if a planet of Jupiter's mass were orbiting the WD
at a distance of 24~AU, which
corresponds to an orbital period of 118~yr,
or a smaller planet in a closer orbit.
Note that reflex motion produces {\it sinusoidal} variations on
the $O-C$, which are only distinguishable from parabolic variations
after a significant portion
of the orbit has been covered. 
As we have observed the star for 25~yr, a sinusoid with a
period shorter than 100~yr can be discarded, but
if the orbiting object were near apoastron in a highly
eccentric orbit, the difference would be harder to distinguish.

\section{Proper Motion}
Pajdosz (1995) discusses the influence of the proper motion of the star
on the measured $\dot P$:
\[\dot P_{\mathrm{obs}} = \dot P_{\mathrm{evol}}\left(1+v_r/c\right)
+ P\dot v_r/c\]
where $v_r$ is the radial velocity of the star. Assuming
$v_r/c  \ll 1$,
he derived
\[\dot P_{\mathrm{pm}} = 2.430 \times 10^{-18} P[s]
\left(\mu["/yr]\right)^2 \left(\pi["]\right)^{-1}\]
where $\dot P_{\mathrm{pm}}$ is the effect of the proper motion
on the rate of period change, $P$ is the pulsation period,
$\mu$ is the proper motion, and $\pi$ is the parallax.
He also calculated that for G117--B15A, 
$\dot P_{\mathrm{pm}} \simeq (8.0 \pm 0.4) \times 10^{-16}$~s/s,
using the proper motion ($\mu=0.136 \pm 0.002~"/{\mathrm{yr}}$)
and parallax $\pi=(0.012 \pm 0.005")$ measured by
Harrington \& Dahn (1980). With the parallax by
Van Altena et al. (1995) of $\pi=(0.0105 \pm 0.004")$, and 
the above proper motion, we calculate
$\dot P_{\mathrm{pm}}= (9.2 \pm 0.5) \times 10^{-16}$~s/s.

The upper limit to the observed $\dot P$ is already only  a few times
the $\dot P$ expected from proper motion alone.

\section{Conclusions}
While it is true that the period change timescale can be proportional to
the cooling timescale, other phenomena with 
shorter timescales can affect $\dot P$. The cooling timescale is the longest
possible one. As a corollary, if the observed $\dot P$ is low enough
to be consistent with evolution, then other processes (such as perhaps a
magnetic field) are not present at a level sufficient to affect $\dot P$.

We compare the observed value
of \Pdot\ with
the range of theoretical values derived from realistic evolutionary models
with $C/O$ cores
subject to {\it g}--mode pulsations
in the temperature range
of G117--B15A.
The adiabatic pulsation calculations
of 
Bradley (1996), and
Brassard \etal\ (1992,1993), which allow for mode trapping, give
\Pdot $\simeq (2-7) \times 10^{-15} \,\rm s/s$
for the $\ell=1$, low {\it k} oscillation observed.
The observed 3$\sigma$ upper limit, 
$\dot P \leq 6.5 \times 10^{-15} \,\rm s/s$, corresponding to
a timescale for period change of
$P/\dot P \leq 1.2 \times 10^9$ yr,
equivalent to 1~s in $6 \times 10^6$~yr, is within the theoretical
predictions and very close to it. 

Our upper limit to the rate of period change brings us to realms
where reflex motion from the proper motion companion,
if they form a physical binary, 
or an unseen orbiting planet
is of the same order as the evolutionary timescale.
The effect of proper motion of the star itself is only
a few times smaller. These two effects must therefore be 
accurately measured.
We are on the
way to {\it measure} the evolutionary time scale
for this lukewarm white dwarf, but the observed phase scatter of the
order of 1.8~s increased the baseline necessary for a measurement.
This scatter is still present in our measurement.

\acknowledgments
This work was partially supported by
grants from CNPq (Brazil), FINEP (Brazil), 
NSF (USA),  NASA (USA).

\begin{table}
\caption{Selected $\dot P$ values derived in the 1990's}
\begin{center}
\begin{tabular}{lr}
1992&$(3.2 \pm 3.0) \times 10^{-15}$\\
1995&$(1.2 \pm 2.9) \times 10^{-15}$\\
1997&$(1.2 \pm 2.2) \times 10^{-15}$\\
1999&$(2.8 \pm 1.7) \times 10^{-15}$\\
2000&$(2.3 \pm 1.4) \times 10^{-15}$\\
\end{tabular}
\end{center}
\end{table}

\begin{figure}
\epsscale{0.7}
\plotone{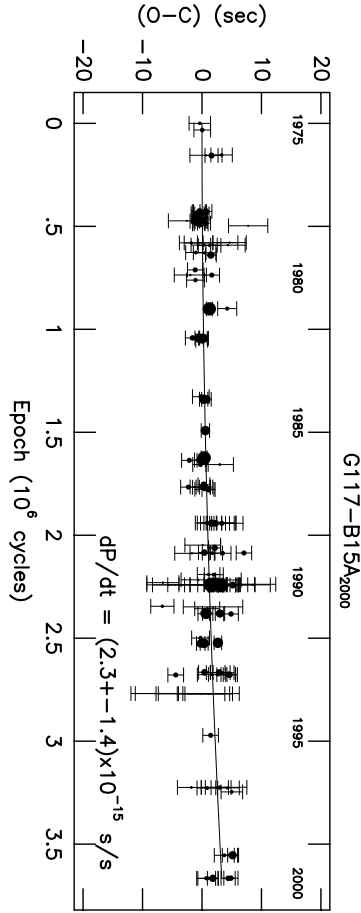}
\figcaption[g117omc.ps]{
{\bf (O-C)}: ({\bf O}bserved minus {\bf C}alculated times of maxima) for
the 215~s pulsation of G117-B15A. The size of each point is proportional
to its weight, i.e., inversely proportional to the uncertainty
in the time of maxima squared. We show 2$\sigma$ error bars for each point,
and the line shows our best fit parabola to the data.
Note that as the period of pulsation is 215.197~s, the whole plot shows
only $\pm 36\deg$ in phase. At the top of the plot, we show the year
of the observation.}
\end{figure}
\end{document}